\documentstyle[11pt,aaspp,astrop-bib,times]{article}
\input{psfig}

\received{19 Dec 1996}
\accepted{15 Jan 1998}

\slugcomment{To appear in PASP (May 1999 issue)}

\lefthead{T. R. Bedding}
\righthead{New methods for speckle interferometry}

\newcommand{\coude}{coud\'{e}}
\newcommand{\about}{\mbox{$\sim$\,}}	
\newcommand{\figref}[1]{Figure~\ref{#1}}
\newcommand{\Figref}[1]{Figure~\ref{#1}}
\begin{document}

\title{New methods for masked-aperture and speckle interferometry}

\author{Timothy R. Bedding}

\affil{School of Physics, University of Sydney 2006, Australia\\ 
{\rm E-mail:} {\tt bedding@physics.usyd.edu.au}}

\begin{abstract} Diffraction-limited images can be obtained with a large
optical telescope using interferometry.  One such method for objects of
sufficient brightness is non-redundant masking (NRM), in which observations
are made through a pupil mask that contains an array of small holes.
However, NRM only uses a small fraction of the available light.  Here I
describe a method for Extended NRM in which a cylindrical lens allows
interferograms from many one-dimensional arrays to be recorded side-by-side
on a two-dimensional detector.  

For fainter objects, the holes in the aperture mask should be replaced by
slits.  In this case, the mask can be removed entirely, with the
cylindrical lens effectively creating a continuous series of
one-dimensional interferograms.  This modified form of speckle
interferometry, which I call MODS (Multiplexed One-Dimensional Speckle), is
intermediate between NRM and conventional full-aperture speckle.  An
existing speckle camera can easily be converted to MODS observations by
inserting a cylindrical lens.  The feasibility of both MODS and Extended
NRM are demonstrated using observations with MAPPIT at the Anglo-Australian
Telescope.
\end{abstract}


\section{Introduction}

There are two approaches to achieving high angular resolution with a large
ground-based telescope.  Both involve compensating for distortions in the
wavefront of the light that result from its passage through the atmosphere.
The first approach is to make these corrections in real time using an
adaptive optics system, which employs a deformable mirror whose shape is
controlled by a large number of actuators (see \citebare{Bec93} for a
review).  However, although adaptive optics shows great promise for
infrared imaging, its application to visible wavelengths poses formidable
problems because of the very large number of actuators required.

The passive approach to high-resolution imaging relies on interferometry
and involves recording many short-exposure images, each of which `freezes'
the atmospheric turbulence.  These images are processed off-line by
calculating the power spectrum and bispectrum, which yield the visibilities
and closure phases of the object.  This technique is known as speckle
imaging \cite{Lab78,Wei91,NegReg96} and can be used to reconstruct a
diffraction-limited image.

A variation on this passive approach is non-redundant masking (NRM), in
which the short-exposure images are taken through a pupil mask that
contains a small number of holes, arranged so that all the baseline vectors
are distinct \cite{HMT87,NKG89}.  Masking the telescope pupil in this way,
although only feasible for bright objects, has several advantages
(\citebare{Han94}; see also \citebare{BvdLZ93}).  These include: (i)~an
improvement in signal-to-noise ratios for the individual visibility and
closure phase measurements; (ii)~attainment of the maximum possible angular
resolution by giving full weight to the longest baselines; and (iii)~a
resistance to variations in atmospheric conditions and a consequent
improvement in the accuracy of visibility calibration.  The
spatial-frequency plane is coarsely sampled relative to observations with a
fully-filled aperture, but each measurement is more accurate.

NRM has been successfully used to image close binaries and to measure
angular diameters of cool stars \cite{HMT87,NKG89,BRM94b,HST95,BZvdL97}.
Most importantly, it has revealed the presence of hotspots and other
asymmetries on the surface of red supergiants and Mira variables
\cite{BHB90,WBB92,HGG92,THB97,BZvdL97}.

\section{Extended NRM}	\label{sec.ENRM}

Two disadvantages of NRM stem from its use of only a small fraction of the
telescope pupil: (i)~the instantaneous coverage of spatial frequencies is
sparse; and (ii)~most of the available light is discarded.  The first point
can be mitigated by combining observations made with different masks and/or
with the masks rotated to several different position angles on the sky.
The second is more serious and is presumably responsible for a reluctance
in the wider community to make use of aperture masks.  Interestingly,
similar considerations have not prevented the use of detectors with low
duty cycles, such as intensified CCDs that are only capable of recording a
few frames per second \cite{WBH96,KEM97}.  In contrast, NRM experiments
have been able to take advantage of fast one-dimensional detectors with
100\% duty cycle, in the form of CCDs with on-chip binning \cite{BHB90}.

In any case, it would clearly be desirable to devise a scheme which makes
use of the whole pupil while maintaining the advantages of NRM\@.
\citeone{Kul88} discusses methods for so-called Extended NRM, in which the
pupil is divided into many slices that are treated separately.  One version
requires an instrument with a series of masks, each transmitting a fraction
of the light to a detector and reflecting the remainder to subsequent
masks.  An alternative method is to image the pupil onto a bundle of
optical fibres, which are again divided among several detectors.

\begin{figure*}

\centerline{
\psfig{%
bbllx=40mm,bblly=104mm,bburx=197mm,bbury=199mm,%
figure=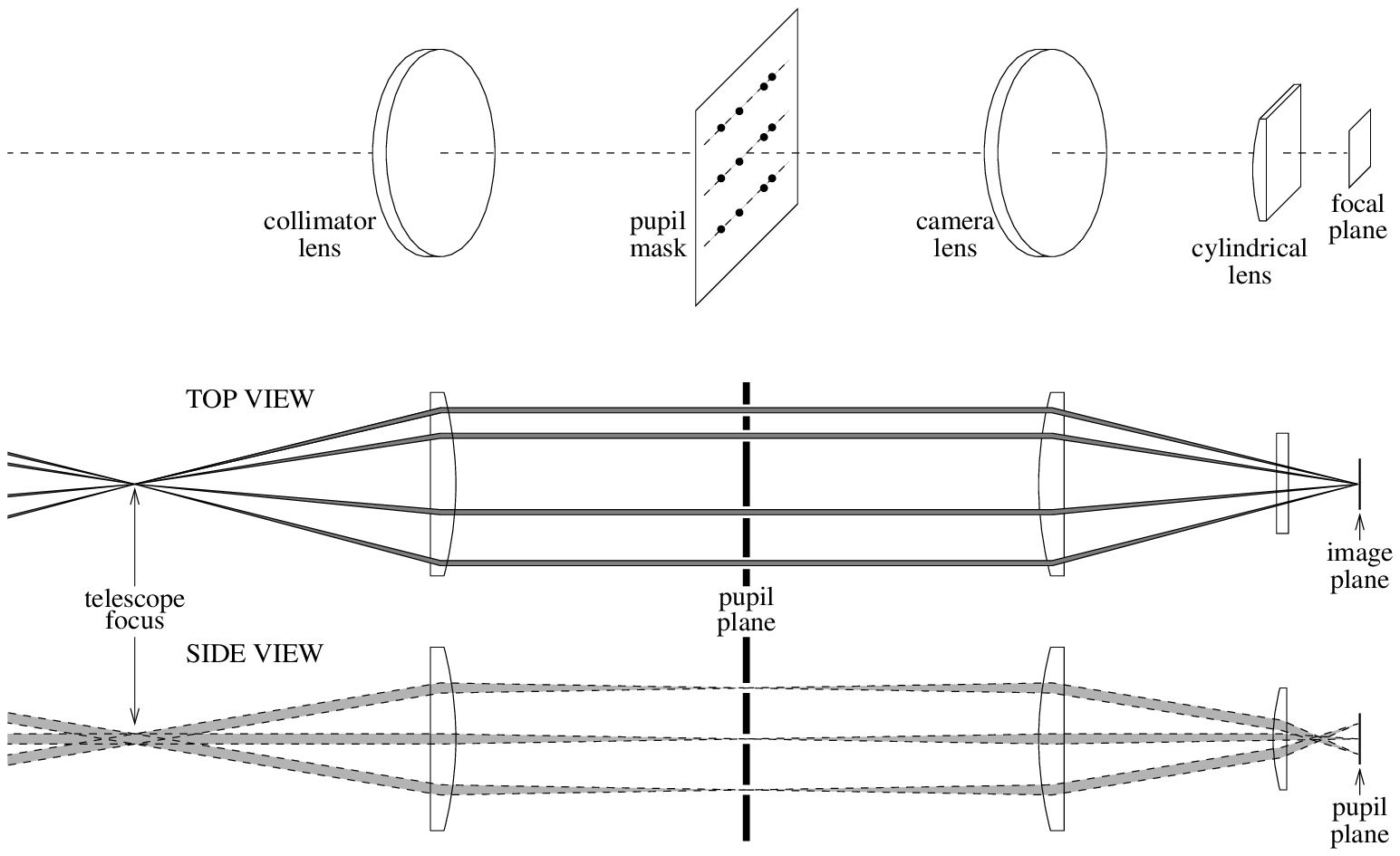,%
width=\the\hsize}}

\caption[]{\label{fig.diagram} Extended NRM using multiple hole arrays and
a cylindrical lens. }

\end{figure*}

A much simpler approach, mentioned briefly by \citeone{Bus88}, is to use a
cylindrical lens.  A method for this is shown schematically in
\figref{fig.diagram}.  The mask, which contains several parallel linear
arrays of holes, is placed in a collimated beam.  The optics in the
interference direction (top view) form a conventional image-plane
interferometer, with the camera lens producing an image of the star crossed
by interference fringes.  In the orthogonal direction (side view), the
cylindrical lens produces a {\em pupil\/} image, which ensures that the
beams from the different hole arrays are spatially separated in the focal
plane.  For simplicity, \figref{fig.diagram} shows a mask with three
identical four-hole arrays.  In practice, more than three arrays would be
used and they could all be different.  Also required, but not shown in the
diagram, are a narrow-band filter and a two-dimensional detector.  A
microscope objective may also be necessary to ensure sufficient
magnification.

Note that the arrangement proposed here is very similar to the
wavelength-dispersed system developed by \citeone{BRM94b}, but with the
dispersing prism being replaced by a narrow-band filter and with a mask
having several parallel arrays of holes.  The second dimension of the
detector is now used, not for wavelength information, but for recording
many simultaneous sets of fringes.

\subsection{Test observations of Extended NRM}

A test of this concept was made at the \coude\ focus of the 3.9-metre
Anglo-Australian Telescope (AAT) using the MAPPIT facility, which was
developed for interferometry and NRM \cite{BRM94b}.  The components of
MAPPIT are mounted on two optical rails attached to the telescope
foundation, providing great freedom to experiment with different optical
configurations.  The tests were made on 1995 January 14 during periods of
intermittent cloud which had interrupted the scheduled observing program.
Despite the cloud, the seeing was quite good (\about1$''$).  The detector
was a $1024\times1024$ Thomson CCD with 19\,\micron\ pixels, although only
a subset of the CCD was read out.  The wavelength region was selected using
an interference filter with central wavelength of 650\,nm and a
transmission bandwidth of 40\,nm.

\begin{figure*}

\centerline{
\psfig{figure=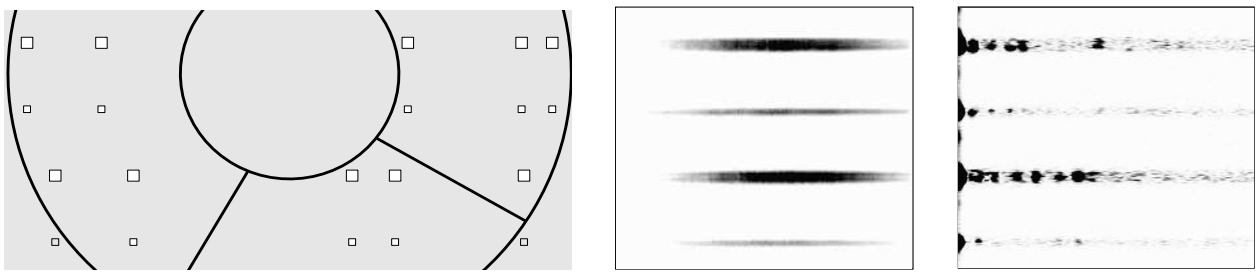,%
bbllx=107pt,bblly=326pt,bburx=506pt,bbury=413pt,width=\the\hsize%
}}

\caption[]{\label{fig.masked} Extended NRM observation of the star Canopus.
Left: the approximate orientation of the mask and AAT pupil (note that the
outer two holes in the bottom array are not illuminated).  Centre: a single
2-second exposure.  Right: the row-by-row power spectrum, with spots
corresponding to the spatial frequencies sampled by the mask.}

\end{figure*}

\Figref{fig.masked} shows an observation of a bright star made with this
system.  The image (centre) is a single 2-second exposure and contains
225$\times$205 pixels.  The horizontal scale is 0.016$''$ per pixel, so the
image width is 3.5$''$ on the sky.  In the vertical direction, which
corresponds to an image of the masked pupil, one row on the CCD projects to
0.9\,cm on the primary mirror.  The diagram on the left shows the
approximate position of the mask with respect to the AAT pupil.  As can be
seen, the mask holes are square and come in two sizes --- these have
projected diameters of 5 and 8\,cm, respectively.

The right panel of \figref{fig.masked} shows the result of calculating the
power spectrum of each row of the image.  Zero spatial frequency is at the
left, and each baseline sampled by the mask produces a spot.  Despite the
relatively long exposure (2\,s), which was substantially longer than the
atmospheric coherence time, power is detected on most baselines in this
single exposure.  In practice, one would use a large number of shorter
exposures.

This preliminary test demonstrates the feasibility of Extended NRM\@.  The
mask was not designed for this application, and two of the holes actually
fall outside the AAT pupil.  An optimized system would use a mask with many
more arrays, perhaps having different numbers of holes with various
distributions and sizes.  Extracting the object visibilities and closure
phases, followed by model fitting or image reconstruction, would proceed as
for conventional NRM\@.

The improvement over conventional NRM comes from the ability to make
observations simultaneously with many parallel mask array.  It should be
possible to fit up to 20--30 different arrays on the pupil, which would
reduce by this factor the total observing time required to acquire a given
amount of data.  However, this gain may be almost completely offset by the
need to use a two-dimensional detector, since one would be forced to a
lower duty cycle than is possible with a fully binned CCD, as are currently
used for conventional NRM \cite{BHB90}.  A practical system based on
Extended NRM should probably wait for a two-dimensional detector with high
duty cycle.

\section{Multiplexed one-dimensional speckle (MODS)}

An obvious way to extend aperture masking to fainter objects is to replace
the array of small holes by a thin slit.  Indeed, a slit was suggested by
\citeone{A+R77} as a good compromise between standard speckle techniques
and Michelson interferometry.  \citeone{B+H93} confirmed this by showing
that the best signal-to-noise for individual visibility and closure phase
measurements at low light levels is achieved with a pupil having a high
degree of redundancy and a small surface area.  This is because the signal
on a given baseline (or closure-phase triangle) increases with the
redundancy, while photon noise increases with the total area of the pupil
(we are assuming that photon noise dominates any contribution from detector
read noise).  A slit, having high redundancy per unit area, is ideal.  It
largely retains two of the advantages of NRM mentioned in the Introduction
(improved signal-to-noise and full resolution), although the accuracy of
visibility calibration in the presence of variable seeing is not much
better than for conventional full-aperture speckle.

The method described in Section~\ref{sec.ENRM} can easily be applied to a
mask with slits.  The cylindrical lens ensures that each slit is imaged
separately on the detector.  Furthermore, we can imagine adding more and
more slits until they become so close together that they fill the whole
pupil.  At this point we can remove the mask entirely.  Such an arrangement
effectively has many adjacent pseudo-slits, and I will refer to it as
Multiplexed One-Dimensional Speckle (MODS).

\subsection{Test observations of MODS}

\begin{figure*}

\centerline{
\psfig{figure=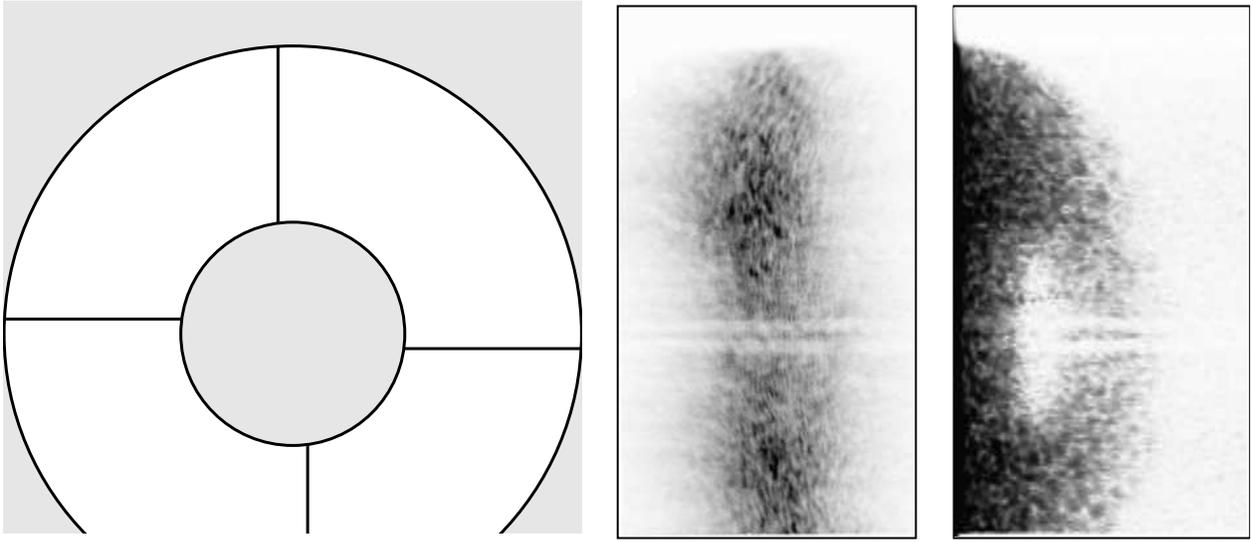,%
bbllx=101pt,bblly=184pt,bburx=514pt,bbury=362pt,width=\the\hsize%
}}

\caption[]{\label{fig.single} MODS observation of the star $\beta$~Cen.
Left: the approximate pupil orientation.  Centre: a single 0.2-second
exposure.  Right: the row-by-row power spectrum averaged over ten such
exposures.  }
\end{figure*}

Observations using the MODS technique were made using the setup described
above, but with the mask removed.  \figref{fig.single} presents
observations of a bright star.  The diagram on the left shows the AAT
pupil, including the obstruction from the secondary mirror and the vanes
that support it.  The image (centre) is a single 0.2-second exposure and
contains 225$\times$400 pixels.  The horizontal and vertical scales are the
same as in \figref{fig.masked}.  Speckle patterns and the shadows of the
vanes are clearly visible, as are fine fringes due to interference across
the central obstruction.

As before, we can process each row of the detector independently.  The
row-by-row power spectrum is shown in the right panel of
\figref{fig.single} and shows power out to the diffraction limit.  The
central obstruction of the AAT is more than one third of the telescope
diameter, so the central part of the row-by-row power spectrum contains a
gap in the coverage of spatial frequencies.

\begin{figure*}

\centerline{
\psfig{figure=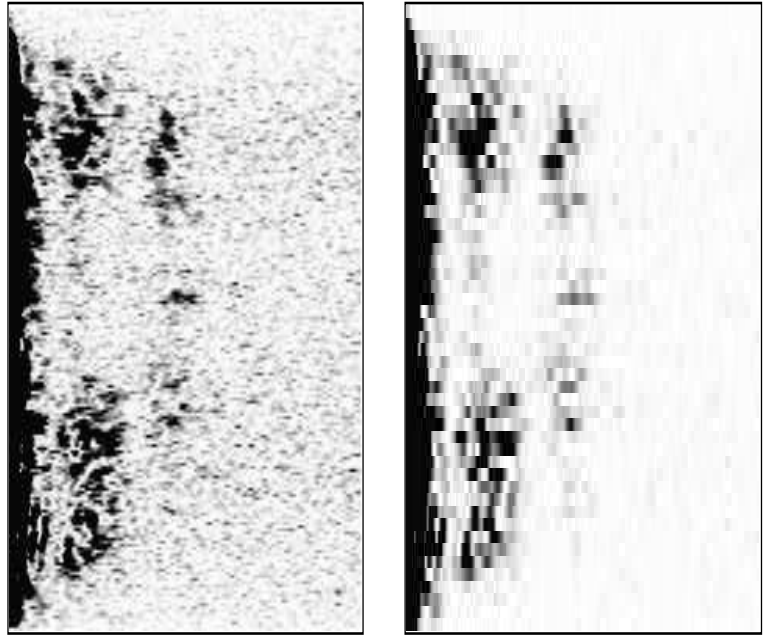,%
bbllx=195pt,bblly=232pt,bburx=409pt,bbury=412pt,width=10cm%
}}

\caption[]{\label{fig.double} MODS observation a double star
($\gamma$~Cen).  Left: the row-by-row power spectrum of a single 0.2-second
exposure.  Right: the same, except the image was binned eight rows at a
time before calculating the power spectrum.  }

\end{figure*}

Observations were also made of the double star $\gamma$~Cen (HR 4819),
which has equal components separated by 1.2$''$ \cite{H+S}.  The position
angle of the observation was chosen so that the separation vector of the
two stars was almost perpendicular to the axis of the cylindrical lens.  In
this way, the projected separation of the binary along the interferometer
direction was only 0.15$''$.  The left panel of \figref{fig.double} shows
the row-by-row power spectrum of a single 0.2-second exposure.  The
vertical stripes are the clear signature of a double star.

The right panel of \figref{fig.double} shows the row-by-row power spectrum
of the same data, but here the original image was binned eight rows at a
time before transforming.  Each row, instead of projecting to 0.9\,cm on
the primary mirror, now projects to 7\,cm.  Thus, by binning several rows
together, we have been able to increase the ``slit'' width in
post-processing and decrease the noise from photon statistics, as can be
seen in the figure.  In practice, the optimum binning factor would depend
on the atmospheric coherence length ($r_0$) at the time of observation and
could be determined during data reduction.  In fact, as shown by
\citeone{Bus88b}, the optimum aperture size for measuring visibilities is
not the same as for closure phases.  With MODS data one would be able to
accommodate this by varying the binning factor in post-processing.

The analysis of MODS data, as with conventional speckle and NRM techniques,
proceeds by estimating the power spectrum and closure phases of the object.
The first step is to calculate the row-by-row power spectrum, as described
above.  Calibration for atmospheric and instrumental effects would be done
using similar observations of an unresolved reference star.  The result
would be a series of measurements of the object power spectrum at many
different spatial frequencies.  Estimates of the closure phase would be
obtained in a similar way by calculating the row-by-row bispectrum.
Together, these visibility amplitudes and closure phases would be used
either to fit a model or to produce and image by deconvolution.  These
processes are now standard in optical interferometry --- the advantage of
the MODS system is the multiplexing of data collection.

\section{Discussion and conclusions}

The observations reported here have shown the feasibility of using a
cylindrical lens for Extended NRM and for MODS\@.  In these test
observations, the detector scale was not optimal in that the vertical
direction was oversampled.  In principle, only one or two pixels are
required across each $r_0$-sized strip on the pupil.  Oversampling
increases the contribution from readout noise and, unlike the case of
photon noise, this cannot be reduced by rebinning in post-processing.  The
best solution would be to use a more powerful cylindrical lens, which would
reduce the height of the pupil image.  Alternatively, the effective pixel
size on the CCD could be increased by on-chip binning.

For MODS observations, no aperture mask is used and so there is no need to
form an intermediate pupil image.  Thus, an existing speckle camera could
be modified to perform MODS merely by inserting a cylindrical lens in front
of the telescope focus.  The lens should be positioned so as to image the
telescope pupil onto the focal plane.  It is not hard to show that the
height of the re-imaged telescope pupil would be equal the focal length of
the cylindrical lens divided by the focal ratio of the telescope beam.  For
example, at the $f/36$ \coude\ focus of the AAT, a cylindrical lens with
$f=100$\,mm would produce a pupil image 2.8\,mm high.

With such a modified speckle camera, measurements at different position
angles could be obtained by rotating the cylindrical lens.  Unless the
detector were also rotated, this would require the detected image to be
de-rotated before one could perform the row-by-row analysis described
above, but such rebinning should not present serious problems.
Reconstructing a two-dimensional image from this series of one-dimensional
measurements is a standard procedure, as described for example by
\citeone{B+H93}.  Since MODS uses the full pupil, the number of detected
photons is the same as for conventional speckle (neglecting any slight
reflective losses from the extra lens).  The advantage is a higher
signal-to-noise on individual visibility and closure phase measurements, at
the expense of requiring observations at several different position angles.

How would the performance of a speckle camera be enhanced by this
modification?  The same number of photons would be collected, so one should
not expect a large change in the limiting magnitude.  Nor would MODS bring
any improved resistance to variations in the seeing.  However, there would
be a gain in angular resolution.  This arises because a one-dimensional
pupil gives more weight to the longer baselines.  It has been demonstrated
that NRM can resolve binaries stars with the AAT down to separations of
15\,mas \cite{RBA99}, whereas full-aperture speckle has only given useful
results with 4-m class telescopes for separations above about 25\,mas.  In
addition, observations using MODS would spread the starlight more uniformly
over the detector, making detector non-linearities much less important.
This is especially useful for detectors that employ an image intensifier,
which are common in speckle cameras, and is seen by \citeone{B+H93} as the
most important advantage of pupil apodization at optical wavelengths.

The main feature of MODS is that it confines the high-resolution
information to one dimension, thereby increasing the contrast of the
speckle pattern and improving the SNR of the visibility and closure phase
measurements.  As discussed by \citeone{B+H93}, this allows measurement of
closure phases in regions of the bispectrum that would otherwise be
useless.  The key advantage of MODS is therefore an ability to image
similar objects to speckle with higher fidelity and dynamic range.  Image
quality is frequently an important issue in extracting science from speckle
images, so there is clearly a potential for the MODS technique to make an
important contribution.

\acknowledgments

The observations could not have been made without the help of Gordon
Robertson, Ralph Marson and John Barton.  I also thank Gordon Robertson and
Ralph Marson for comments on this paper.  The development of MAPPIT was
supported by a grant under the CSIRO Collaborative Program in Information
Technology, and by funds from the University of Sydney Research Grants
Scheme and the Australian Research Council.

\end{document}